\documentclass[reprint,  superscriptaddress]{revtex4-1}

\usepackage{amsmath,amssymb,amsfonts}
\usepackage{mathrsfs,mathtools} 

\usepackage[T1]{fontenc}

\usepackage{graphicx}
\usepackage{subcaption}

\usepackage[ruled]{algorithm2e}

\usepackage[braket]{qcircuit}

\usepackage{dcolumn}
\usepackage{tabularx}
\usepackage{array}
\newcolumntype{Z}{>{\raggedright\let\newline\\\arraybackslash\hspace{0pt}}X}

\begin{document}

\title{Protocols for creating and distilling multipartite GHZ states with Bell pairs}

\author{Sébastian de Bone}
\affiliation{QuTech, Delft University of Technology, Lorentzweg 1, 2628 CJ Delft, The Netherlands}
\affiliation{QuSoft, CWI, Science Park 123, 1098 XG Amsterdam, The Netherlands}
\author{Runsheng Ouyang}
\affiliation{Department of Physics, Tsinghua University, Beijing 100084, China}
\affiliation{Department of Information Technology and Electrical Engineering, ETH Zürich, Gloriastrasse 35, 8092 Zürich, Switzerland}
\author{Kenneth Goodenough}
\affiliation{QuTech, Delft University of Technology, Lorentzweg 1, 2628 CJ Delft, The Netherlands}
\author{David Elkouss}
\email{d.elkousscoronas@tudelft.nl}
\affiliation{QuTech, Delft University of Technology, Lorentzweg 1, 2628 CJ Delft, The Netherlands}

\begin{abstract}
The distribution of high-quality Greenberger-Horne-Zeilinger (GHZ) states is at the heart of many quantum communication tasks, ranging from extending the baseline of telescopes to secret sharing. They also play an important role in error-correction architectures for distributed quantum computation, where Bell pairs can be leveraged to create an entangled network of quantum computers. We investigate the creation and distillation of GHZ states out of non-perfect Bell pairs over quantum networks. In particular, we introduce a heuristic dynamic programming algorithm to optimize over a large class of protocols that create and purify GHZ states. All protocols considered use a common framework based on measurements of non-local stabilizer operators of the target state (\textit{i.e.}, the GHZ state), where each non-local measurement consumes another (non-perfect) entangled state as a resource. The new protocols outperform previous proposals for scenarios without decoherence and local gate noise. Furthermore, the algorithms can be applied for finding protocols for any number of parties and any number of entangled pairs involved.
\end{abstract}

\maketitle

\section{Introduction}
Quantum computation promises a computational advantage for algorithmic problems in the fields of cryptography, database searching, simulations of atoms and molecules, and solving linear equations. There are several approaches and technologies concurrently investigated for scaling the near term quantum devices to full-fledged quantum computers. One of the approaches is \textit{distributed} or \textit{networked} quantum computing \cite{ciracDistributedQuantumComputation1999,groverQuantumTelecomputation1997}. In this approach, multiple computers holding a small number of qubits are connected via entanglement~\cite{vanmeterPathScalableDistributed2016}.

The promise of distributed quantum computing is the possibility of building a quantum computer without the difficulty of engineering a large multi-qubit device. In exchange, the feasibility of such a networked device critically lies in the availability of high-fidelity entanglement. This is because entanglement is required for the realization of multi-qubit operations between different quantum computers. In particular, entangled states are necessary for performing error detection measurements in error-correction codes executed with distributed quantum computers. 

Using error-correction for fault-tolerant quantum computation relies on encoded data. To correct or track the errors on the encoded data, it is necessary to periodically perform joint measurements on different qubits. If the whole encoded state lies in a single quantum device, these joint measurements can be performed by applying the appropriate multi-qubit operations and measuring an ancilla qubit. However, in distributed implementations the joint measurements become non-local. The ingredient that enables the joint measurements are \textit{Greenberger-Horne-Zeilinger} (GHZ) states. By consuming an $n$-qubit GHZ state it is possible to perform a non-local measurement between $n$ parties. The challenge of distributed quantum computation is to produce GHZ states at a fast enough rate and with high enough fidelity to enable fault-tolerant quantum computation. 

Creating GHZ states is experimentally challenging. A simple protocol for creating an $n$-qubit GHZ state consists of \textit{fusing} $n-1$ Bell pairs. However, the fidelity of the GHZ state degrades exponentially with $n$. This problem can be overcome by more complicated protocols that \textit{distill} or \textit{purify} either the input Bell states or any of the intermediate states of a protocol. This generally improves the fidelity of the final GHZ state, but comes at the price of consuming a larger number of Bell pairs.

Several physical systems can process quantum information and have a coherent optical interface for generating remote entanglement \cite{awschalomQuantumTechnologiesOptically2018}. Some examples are nitrogen-vacancy (NV) centres \cite{Taminiau2014, Cramer2016, reisererRobustQuantumNetworkMemory2016, Kalb2017a, Bradley2019, Abobeih2019}, silicon-vacancy (SiV) centres \cite{Sipahigil2016, sukachevSiliconVacancySpinQubit2017, nguyenQuantumNetworkNodes2019, nguyenIntegratedNanophotonicQuantum2019}, and ion traps \cite{huculModularEntanglementAtomic2015, Nigmatullin2016}. Some of these platforms have already demonstrated the generation of long lived remote entanglement \cite{humphreysDeterministicDeliveryRemote2018} and even distillation \cite{Kalb2017a}. However, the rate at which entanglement can be produced is slower than the gate times. In consequence, the rate at which GHZ states are produced becomes the bottleneck for the performance of distributed quantum computer implementations \cite{nickersonFreelyScalableQuantum2014}. Moreover, while our motivation stems from distributed quantum computation, efficient GHZ generation has direct application in several other applications including secret sharing \cite{hilleryQuantumSecretSharing1999}, anonymous transmission \cite{christandlQuantumAnonymousTransmissions2005}, clock synchronization \cite{komarQuantumNetworkClocks2014}, and extending the baseline of telescopes \cite{khabiboullineOpticalInterferometryQuantum2019}.

The goal of our research is to minimize the number of Bell pairs necessary to produce high-fidelity GHZ states. We do this by searching the protocol space for creating GHZ states out of Bell pairs. The difficulty of the problem is that given a number of parties and a number of input Bell pairs, the number of possible protocols is very large. In fact, it grows super-exponentially with these parameters. Our approach to deal with the large number of protocols is therefore to take the heuristic approximation that optimal protocols for some number of copies of a GHZ state are composed of optimal protocols for a smaller number of copies or parties. This heuristic leads to a \textit{dynamic program}. 

Distillation is better understood in the bipartite case~\cite{Bennett1996,Deutsch1996,fujiiEntanglementPurificationDouble2009,krastanovOptimizedEntanglementPurification2019} than in the multipartite case~\cite{Murao1998, maneva2002improved, Dur2003, hoPurifyingGreenbergerHorneZeilingerStates2008, glancyEntanglementPurificationAny2006, kruszynskaEntanglementPurificationProtocols2006, huberPurificationGenuineMultipartite2011, hostensHashingProtocolDistilling2006, hostensStabilizerStateBreeding2006}. In the bipartite case, it is even known that some protocols achieve an optimal trade-off between rate and fidelity \cite{Rozpedek2018}. 

In the context of a distributed implementation of the surface code, Nickerson \textit{et al.} \cite{Nickerson2013a} optimized a family of protocols for generating four-partite GHZ states out of noisy Bell pairs. To facilitate experimental feasibility, the GHZ distillation protocols require three qubits per node. The number of possible protocols in this family, while large, is still brute-force tractable. Subsequent work optimized a similar family of protocols in the presence of loss \cite{nickersonFreelyScalableQuantum2014}. We leave the extension of our approach to more realistic settings including loss for future work. In contrast with \cite{Nickerson2013a}, we are interested in more general protocols that minimize the number of Bell pairs consumed independently of the size of the required quantum register. This different ansatz is justified by recent experimental progress with multi-qubit registers \cite{Bradley2019}. 

In sections \ref{sec:diagonal_states} and \ref{sec:operations}, we introduce the formalism and building blocks of the GHZ generation protocols considered. In section \ref{sec:comparison-other-prots}, we show that existing GHZ generation protocols are included in our search space. In section \ref{sec:dynamicprograms}, we present our dynamic program. In section \ref{sec:results}, we show the performance of the best GHZ creation protocols founds. Finally, we draw our conclusions in section \ref{sec:conclusions}.

\section{Bell and GHZ diagonal states} \label{sec:diagonal_states}
Here, we introduce notation and definitions used in the rest of the paper together with our model for states.

We describe non-perfect Bell and GHZ states in the \textit{stabilizer} formalism~\cite{Nielsen2000}. A \textit{stabilizer operator} or \textit{stabilizer} of a quantum state $\ket{\psi}$ is an operator $O$ that verifies $O\ket{\psi}=\ket{\psi}$, i.e. $\ket{\psi}$ is an eigenvector of $O$ with eigenvalue $+1$ and, in consequence, leaves $\ket{\psi}$ invariant . An $n$-qubit pure quantum state has $2^n$ stabilizer operators. These $2^n$ operators form the \textit{stabilizer group} of the state which is generated by a subset of $n$ operators. An $n$-qubit GHZ state $(\ket{0}^{\otimes n}+\ket{1}^{\otimes n})/\sqrt{2}$ is described by the stabilizer group generated by the operators $\{X_1 X_2 \dots X_n, Z_1 Z_2, Z_2 Z_3, \dots, Z_{n-1} Z_n\}$. The stabilizer group includes the identity $\mathbb{I}$. We call the $2^n-1$ operators in this group that are not the identity $\mathbb{I}$ the \textit{non-trivial} stabilizers of the state.

We use the stabilizer formalism to define a basis for a general $n$-qubit system. In analogy with the Bell basis, we call this basis the \textit{GHZ basis}. It is also known as the \textit{cat basis} \cite{maneva2002improved}. The basis states of the $n$-qubit GHZ basis are defined as the $2^n$ states $\ket{\phi^{s_1 s_2 s_3 \dots s_n}}$ with stabilizer generators $s_1 X_1 X_2 \dots X_n$, $s_2 Z_1 Z_2$, $s_3 Z_2 Z_3$, $\dots$, $s_n Z_{n-1} Z_n$, where $s_i\in \{+1, -1\}$ for all $i\in\{1,2,\dots,n\}$. As an example, we show the eight basis states of the 3-qubit GHZ basis in Table \ref{tab:three-qubit-ghz}. For any $n$, the basis state $\ket{\phi^{++\dots+}}$ is the $n$-qubit GHZ state $(\ket{0}^{\otimes n}+\ket{1}^{\otimes n})/\sqrt{2}$. We use the capital $\Phi$ symbol to denote the density matrix corresponding to a basis state---\textit{i.e.}, for a general state in the $n$-qubit GHZ basis $\Phi^{s_1 s_2 \dots s_n}\equiv\ket{\phi^{s_1 s_2 \dots s_n}}\bra{\phi^{s_1 s_2 \dots s_n}}$. 

\begin{table}[]
\begin{tabular}{l|c|ccc}
                    & Computational basis              & $X_1 X_2 X_3$ & $Z_1 Z_2$ & $Z_2 Z_3$ \\ \hline
                    &                                  &               &           &           \\[-0.8em]
 $\ket{\phi^{+++}}$ & $(\ket{000}+\ket{111})/\sqrt{2}$ & $+1$          &  $+1$     &  $+1$     \\ [0.5em]
 $\ket{\phi^{++-}}$ & $(\ket{001}+\ket{110})/\sqrt{2}$ & $+1$          &  $+1$     &  $-1$     \\ [0.5em]
 $\ket{\phi^{+-+}}$ & $(\ket{011}+\ket{100})/\sqrt{2}$ & $+1$          &  $-1$     &  $+1$     \\ [0.5em]
 $\ket{\phi^{+--}}$ & $(\ket{010}+\ket{101})/\sqrt{2}$ & $+1$          &  $-1$     &  $-1$     \\ [0.5em]
 $\ket{\phi^{-++}}$ & $(\ket{000}-\ket{111})/\sqrt{2}$ & $-1$          &  $+1$     &  $+1$     \\ [0.5em]
 $\ket{\phi^{-+-}}$ & $(\ket{001}-\ket{110})/\sqrt{2}$ & $-1$          &  $+1$     &  $-1$     \\ [0.5em]
 $\ket{\phi^{--+}}$ & $(\ket{011}-\ket{100})/\sqrt{2}$ & $-1$          &  $-1$     &  $+1$     \\ [0.5em]
 $\ket{\phi^{---}}$ & $(\ket{010}-\ket{101})/\sqrt{2}$ & $-1$          &  $-1$     &  $-1$ 
\end{tabular}
\caption{Each column of the table shows from left to right: the basis states $\ket{\phi^{s_1 s_2 s_3}}$ of the 3-qubit GHZ basis, their representations in the computational basis and the stabilizers of the state. The signs $s_1$, $s_2$ and $s_3$ describe the relation of the basis state $\ket{\phi^{s_1 s_2 s_3}}$ with stabilizer generators $s_1 X_1 X_2 X_3$, $s_2 Z_1 Z_2$ and $s_3 Z_2 Z_3$. } 
\label{tab:three-qubit-ghz}
\end{table}

For $n=2$, the GHZ basis reduces to the Bell basis. We restrict our attention to states that do not contain off-diagonal elements in the Bell basis which we call \textit{Bell diagonal states}. This restriction does not reduce the applicability of our methods, because any bipartite qubit state can be transformed to a Bell diagonal state with the same fidelity via \textit{twirling} \cite{Bennett1996b}, a procedure that relies on local operations and classical communication.

In the general case, we also restrict our attention to diagonal states in the GHZ basis. This is justified because the operations introduced in section \ref{sec:operations} take GHZ diagonal states to GHZ diagonal states. Therefore, to track the state of a distillation protocol with input Bell diagonal states and composed of these operations it is sufficient to consider GHZ diagonal states. For $n$ parties $p_1$, $p_2$, $\dots$, $p_n$ we write these states as:
\begin{equation}
\rho_{p_1 p_2 \dots p_n}=\quad\sum_{\mathclap{(s_1,s_2,\dots,s_n)\in\{+1,-1\}^n}}\quad A_{s_1 s_2 \dots s_n}\Phi^{s_1 s_2 \dots s_n}.
\label{eq:diagonal_state}
\end{equation}

Unless otherwise stated, in the remainder we use the term \textit{state} to denote both Bell diagonal states and GHZ diagonal states, and call the $A_{s_1 s_2 \dots s_n}$ elements in Eq. \ref{eq:diagonal_state} the \textit{coefficients} of the state. The $A_{++\dots +}$ coefficient denotes the \textit{fidelity} of the state with respect to the state $\ket{\phi^{++\dots+}}$. Moreover, we use the shorthand $\ket{\phi_n^+}$  for $\ket{\phi^{++\dots+}}$ when we need to make explicit the number of qubits of the state. Finally, we let $F_\text{Bell}$ be the fidelity of a Bell diagonal state $\ket{\phi_2^+}$ and $F_\text{GHZ}$ or $F_\text{GHZ}^{(n)}$ be the fidelity of an $n$-qubit GHZ diagonal state $\ket{\phi_n^+}$. 
\section{Operations on Bell and GHZ diagonal states} \label{sec:operations}

This section discusses two operations on Bell and GHZ diagonal states: fusion operations, that merge two states, and distillation operations, that consume one state to improve the fidelity of another state. 

\subsection{Fusion} \label{sec:fusion_op}
\begin{figure*}
\centering
\includegraphics[width=0.94\textwidth]{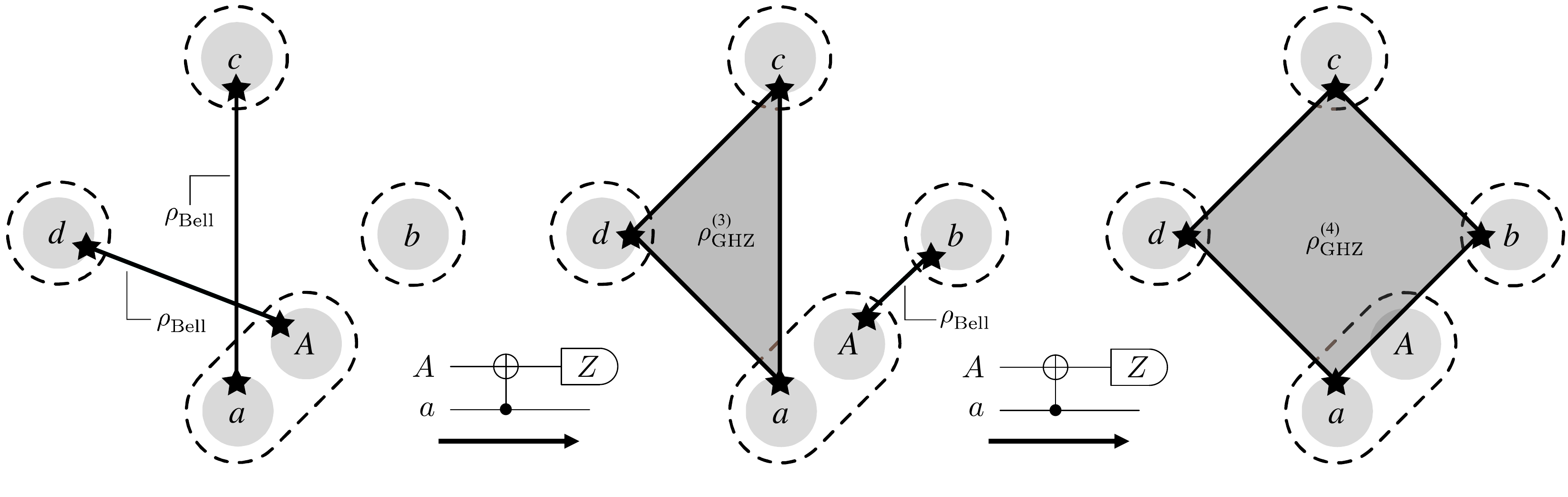}
\caption{The fusion operation allows to merge Bell diagonal states and GHZ diagonal states that overlap at one of the network parties. This party applies the operation depicted on top of the arrow, followed by local Pauli gate corrections that depend on the measurement outcome (not depicted).\label{fig:fusion}}
\end{figure*}

The \textit{fusion} operation involves two states that are fused or merged. The operation takes an $n_1$-qubit state and an $n_2$-qubit state that overlap in one network node in the sense that the node holds (at least) one qubit of each state. The fusion operation consists of a CNOT gate between one qubit of each state, a measurement in the $Z$ basis of one of the two qubits (see Figure \ref{fig:fusion}) and local Pauli gate corrections to the qubits of the other state. If the two qubits involved are qubit $i$ of the $n_1$-qubit state and qubit $j$ of the $n_2$-qubit we say that we are fusing the $n_1$-qubit state at qubit $i$ with the $n_2$-qubit state at qubit $j$. This results in a new $(n_1+n_2-1)$-qubit state. The fusion operation is deterministic.

\subsection{Non-local stabilizer measurements} \label{sec:stabilizer_op}

A non-local stabilizer measurement also involves two states: a \textit{main state} and an \textit{ancillary state}. The ancillary state is consumed to measure a stabilizer operator. 

In the context of a distillation scheme, the stabilizer operator is one of the stabilizers of some target state. Then, the non-local stabilizer measurement can be understood as an error-detection scheme. A $+1$ outcome projects the state into the corresponding eigenspace which is compatible with the target state, while a $-1$ outcome projects the state into the corresponding eigenspace which is orthogonal to the target state. For this reason, the state is kept when the measurement outcome is $+1$ and discarded otherwise.

\begin{figure}
\centering 
\[
  \Qcircuit @C=.7em @R=.7em {
     {p_1} & &\gate{P_1} &\qw  & \qw  & \qw    &\qw  &\qw  & \qw && \\ 
     {p_2} &  &\qw &\gate{P_2} & \qw & \qw        &\qw  &\qw& \qw && \\  
     \vdots    &&&&\ddots&&&&& \\ 
&&&&&&&&&&\\ 
     {p_n} &   &\qw  & \qw  &   \qw &  \qw & \gate{P_n}    & \qw & \qw && \\ 
        &&&&&&&&&\\ 
       {p_1} &   &\ctrl{-6}     &\qw &  \qw 	          & \qw &\qw 	& \measureD{X} & \cw & \rstick{m_1}   \\ 
       {p_2} &   &\qw   &\ctrl{-6} &  \qw 	          & \qw &\qw 	& \measureD{X} & \cw & \rstick{m_2}   \\ {} 
         \vdots    &&&&\ddots&&&\vdots&&&\vdots \\ 
            &&&&&&&&&\\ 
      {p_n} &      &\qw  &\qw &  \qw 	          &\qw  & \ctrl{-6}  	& \measureD{X} & \cw &  \rstick{m_n}       
      \inputgroupv{1}{5}{2em}{3em}{\hspace{-2em}\rho_\text{m}}
      \inputgroupv{7}{11}{2em}{3.5em}{\hspace{-3.5em}\ket{\phi^{++\dots+}}_\text{a}}
} 
\]
\caption{Quantum circuit to measure the non-local $n$-qubit Pauli operator $P_1 P_2 \dots P_n$ on a general $n$-qubit quantum state $\rho_\text{m}$ (the \textit{main state}). The $n$-qubit state $\rho_\text{m}$ is distributed over $n$ parties $p_1, p_2, \dots, p_n$ in such a way that every party has one qubit of $\rho_\text{m}$. The $n$-qubit state $\ket{\phi^{++\dots+}}_\text{a}$ (the \textit{ancillary state}), shared over the same $n$ network parties, enables measuring $P_1 P_2 \dots P_n$  on $\rho_\text{m}$. The parties find the measurement outcome $m=\prod_{i=1}^n m_i$ combining their individual measurements $(m_1, m_2, \dots, m_n)\in\{+1,-1\}^n$.}
\label{fig:non-loc-Pauli-meas}
\end{figure}
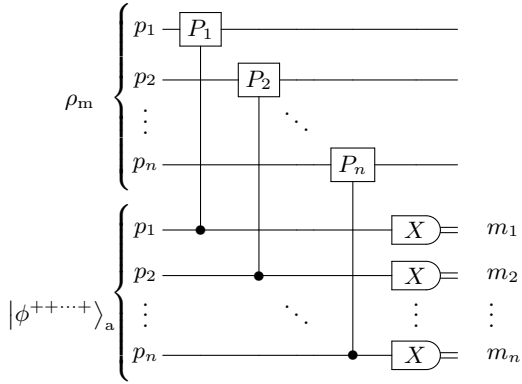

In Fig. \ref{fig:non-loc-Pauli-meas}, we see a quantum circuit that measures a joint Pauli operator $P_1 P_2  \dots P_n$ with the aid of an $n$-qubit state $\ket{\phi^{++\dots+}}=(\ket{0}^{\otimes n}+\ket{1}^{\otimes n})/\sqrt{2}$. The qubits of the ancillary state are measured out individually in the $X$ basis, and the network parties use classical communication to calculate the full measurement outcome. 
While the non-local measurement of the stabilizer in Fig. \ref{fig:non-loc-Pauli-meas} is perfect, in practical situations the ancillary state is noisy and the operation is only carried out approximately.

The $n$-qubit GHZ state $\ket{\phi^{++\dots +}}$ has three different type of stabilizers: $ZZ$ stabilizers of weight 2, combinations of $ZZ$ operators of weight $m$ (where $m\leq n$ is always an even number), and operators of weight $n$ consisting of combinations of $X_1 X_2 \dots X_n$ and any number of the $ZZ$ operators. Measuring the latter type non-locally requires another $n$-qubit ancillary state. The $ZZ$ operators can be measured with a Bell  state as ancillary state. For combinations of $ZZ$ operators of higher weight $m\leq n$, a GHZ state of weight $m$ is required.

\section{GHZ distillation protocols} \label{sec:comparison-other-prots}
Many GHZ creation protocols can be described by combining the operations from section \ref{sec:operations}. In particular, fusion operations create larger multipartite states and non-local stabilizer measurements can be used to increase the fidelity of the main state.  As an example, in the following, we describe two protocols from Nickerson \textit{et al.} \cite{Nickerson2013a} for creating and distilling a GHZ state shared by $n=4$ network parties: one that uses $k=22$ Bell states (\textit{Expedient}), and one that uses $k=42$ Bell states (\textit{Stringent}). 

\begin{figure}[]
\includegraphics[width=8.5cm]{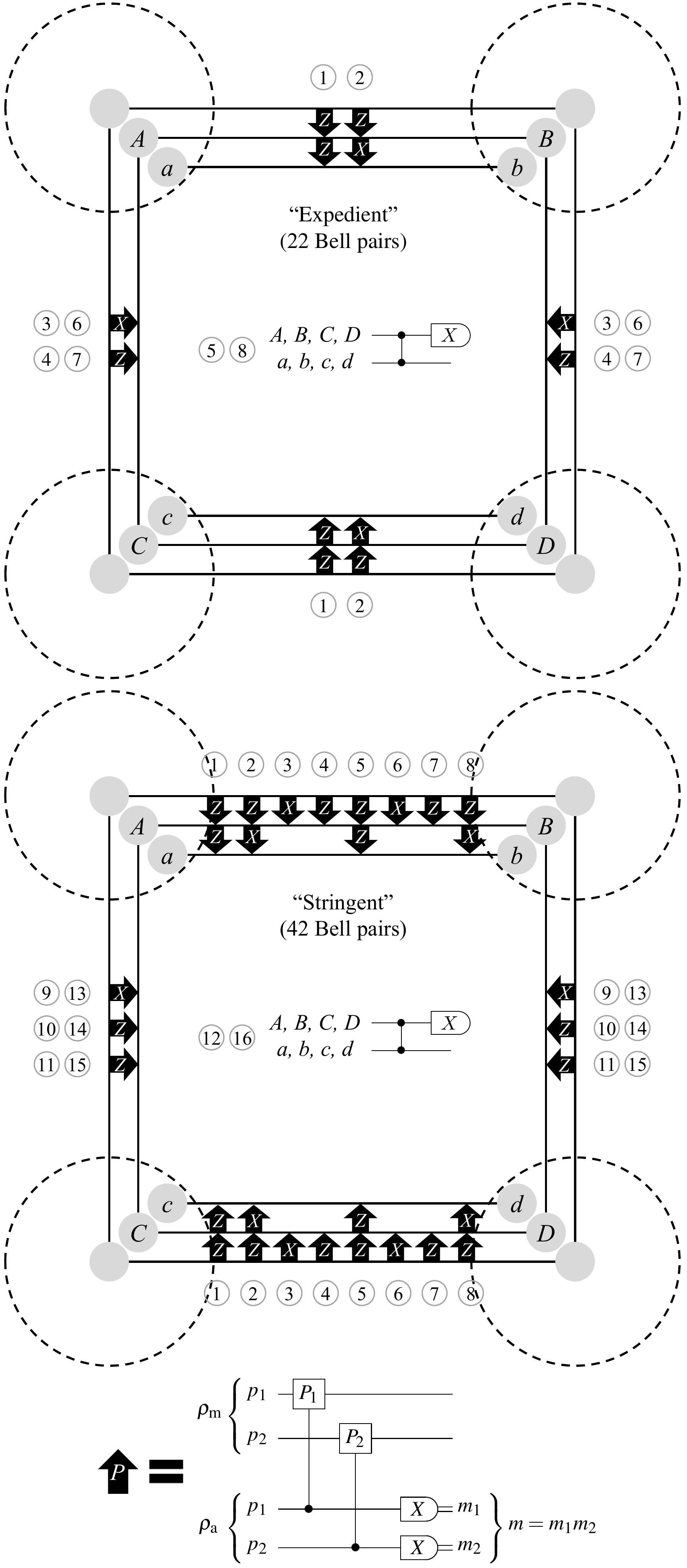}
\centering
\caption{Two protocols for creating and purifying a 4-qubit GHZ diagonal state out of Bell diagonal states shared between 4 network parties~\cite{Nickerson2013a}. For both protocols, each node (dotted circles) requires three qubits (solid grey dots). The numbers denote the order in which individual steps are carried out. Solid lines depict Bell pairs shared between qubits. Black arrows denote the consumption of a Bell pair---the Bell pairs at the beginning of the arrow need to be regenerated for the next step.} 
\label{fig:Expedient_Stringent}
\end{figure}

Expedient and Stringent are the result of a brute-force search over protocols following a concrete multi-step structure \cite{Nickerson2013a}. The first step consists of non-local measurements involving all qubits on two opposite sides of the network (of the type shown between qubits $a$ and $b$, and between qubits $c$ and $d$ in Fig. \ref{fig:Expedient_Stringent}). After that, the two purified Bell pairs are stored. In the second step, the protocols continue with several rounds of non-local measurements that increase the fidelity of the two Bell pairs in the other direction (between qubits $B$ and $C$, and between qubits $A$ and $D$ in Fig. \ref{fig:Expedient_Stringent}). Then the four Bell pairs fused in the third step into a 4-qubit GHZ state. The fourth step is analogous to the second one, two Bell pairs are purified between qubits $B$ and $C$, and between qubits $A$ and $D$. Finally, the purified Bell pairs are used to perform two $ZZ$ non-local measurements to the 4-qubit GHZ state. 

Fig. \ref{fig:Expedient_Stringent} shows a schematic representation of both protocols. The Expedient protocol is on the left of Fig. \ref{fig:Expedient_Stringent} and consumes 22 Bell states. Steps 1 and 2 create purified Bell pairs between qubits $a$ and $b$, and between qubits $c$ and $d$. Steps 3 and 4 create purified Bell pairs between qubits $B$ and $C$, and between qubits $A$ and $D$. In step 5, the GHZ state is created from these Bell pairs. Steps 6 and 7 are repetitions of steps 3 and 4, and step 8 consists of two $ZZ$ non-local measurements on the GHZ state. 

The Stringent protocol has the same main structure as the Expedient protocol, but consumes 42 Bell states. The protocol is depicted at the bottom of Fig. \ref{fig:Expedient_Stringent}. In comparison with Expedient, Stringent consumes an additional number of Bell states to increase the fidelity of the states that take part in the fusion step; analogously a larger number of Bell states is consumed to create the states to perform the final $ZZ$ non-local measurements. 

\section{Dynamic programs to optimize GHZ generation} \label{sec:dynamicprograms}
In this section, we discuss several algorithms to optimize GHZ generation. In particular, we consider the problem of probabilistically generating an $n$-qubit GHZ state with maximum fidelity $F_{\text{GHZ}}$ starting from $k$ isotropic Bell pairs with fidelity $F_{\text{Bell}}$:
\begin{equation}
\rho_{AB}=F_\text{Bell}\Phi^{++}+\quad\sum_{\mathclap{(s_1,s_2)\neq(+,+)}}\quad\frac{1-F_\text{Bell}}{3}\Phi^{s_1 s_2}.
\end{equation}
We do not restrict the number of qubits that the $n$ nodes need to hold, and we also do not restrict their connectivity. 

Ideally, we would consider all possible ways of distributing the Bell pairs over the network parties, and all possible combinations of the operations from section \ref{sec:operations}. Unfortunately, the number of these combinations grows super-exponentially in $n$ (see \cite{goodenoughOptimisingRepeaterSchemes2020} for a similar argument). This makes a brute force approach infeasible for relevant values of $(n,k)$, in particular for the protocols described in Section \ref{sec:comparison-other-prots}: $(n,k)=(4,22)$ and $(n,k)=(4,42)$. For this reason, similar to the approaches in \cite{jiangOptimalApproachQuantum2007, goodenoughOptimisingRepeaterSchemes2020} for Bell pair distribution in the context of quantum repeater chains, we propose heuristic \textit{dynamic programs} for optimizing the distribution of GHZ states. 

The dynamic programs reduce the complexity of the optimization and enable finding good protocols for large values of $n$ and $k$. However, the output is not necessarily optimal. In the following, we first describe a simple dynamic program for optimizing GHZ generation. Next, we build on this description to present a randomized version of the dynamic program. 

\subsection{Base dynamic program}
The dynamic program takes as input the problem parameters $n_{\max},k_{\max},F_{\text{Bell}}$ and a buffer size $b$.
The parameters specify the size of the final $n$-qubit GHZ state, the number of Bell pairs $k$, their fidelity $F_{\text{Bell}}$ and the number of protocols $b$ to store at each intermediate step of the algorithm---\textit{i.e.}, for each combination of the number of Bell pairs and GHZ state size. The pseudo-code of this algorithm can be found in Algorithm \ref{alg:dynamic program}.
 
The algorithm begins with $(n,k)=(2,1)$ and proceeds iteratively combining the solutions for smaller values of $n$ and $k$ until $(n,k)=(n_{\max},k_{\max})$. More precisely, for each value of $(n,k)$ the algorithm combines the protocols found for smaller values of $n,k$ in all possible ways to perform either a non-local measurement or to fuse the states, evaluates the fidelity of the resulting state for each combination and stores the $b$ protocols that achieve the largest fidelity. For $(n,k)=(2,1)$, the algorithm stores a bell pair with fidelity $F_{\text{Bell}}$. The algorithm increases $k$ to $(n,k)=(2,1)$ and it continues increasing $k$, until $(n,k)=(2,k_{\max})$. Then $n$ is increased and $k$ is reset to $n-1$.

\begin{algorithm}[h]
	\caption{Base dynamic program to optimize GHZ generation.}
	\label{alg:dynamic program}
	\SetKwInOut{KwInput}{Input}
    \SetKwInOut{KwOutput}{Output}
	\SetAlgoLined
	\KwInput{$n_\text{max}$: number of qubits of final GHZ state\\
		$k_\text{max}$: number of isotropic Bell pairs\\
		$F_\text{Bell}$: fidelity of isotropic Bell pairs\\
		$b$: number of protocols to store per step
		}
	\KwOutput{Protocol to create an $n$-qubit GHZ state out of $k$ isotropic Bell pairs.}
	\For{$\{(n, k)\, |\, n\leqslant n_\text{max}, \, k\leqslant k_\text{max}, \, k\geqslant n-1\}$}{
		\text{\# Try all non-local measurement combinations.}\\
		\For{stabilizer $\in$ \{stabilizers of $\ket{\phi_n^+}$ \} }{
			\text{$n'$ $\leftarrow$ weight of \textit{stabilizer}}\\
			\For{$k' \in [n'-1, \, k-n+1]$}{
				\text{Measure \textit{stabilizer} on stored $(n,k-k')$ state}\\
				\text{consuming the stored $(n',k')$ state}\\
				\text{Store new state if it has higher fidelity than}\\
				\text{one the existing $(n,k)$ states.}\\ 
				\text{If more than $b$ $(n,k)$ states are stored,}\\ 
				\text{remove the one with lowest fidelity.}
			}
		}
		\text{\# Try all fusion combinations.}\\
		\For{$n_2 \in [2, n-1]$}{
			\text{$n_1\leftarrow n-n_2+1$}\\
			\For{$k_2 \in [n_2 -1, k-n+1]$}{
				\text{$k_1\leftarrow k-k_2$}\\
				\For{$(i, j)\in [0, n_1-1]\times [0, n_2-1]$}{
    				\text{Fuse stored states $(n_1,k_1)$ at qubit $i$ and}\\
    				\text{$(n_2,k_2)$ at qubit $j$}\\ 
    				\text{Store new state if it has higher fidelity}\\
	    			\text{than the existing $(n,k)$ state}\\
	    			\text{If more than $b$ $(n,k)$ states are stored,}\\ 
				    \text{remove the one with lowest fidelity.}
				}
			}
		}
	}
	\Return{stored $(n_\text{max},k_\text{max})$ state with highest fidelity}
\end{algorithm}

\begin{figure}[]
\includegraphics[width=\linewidth]{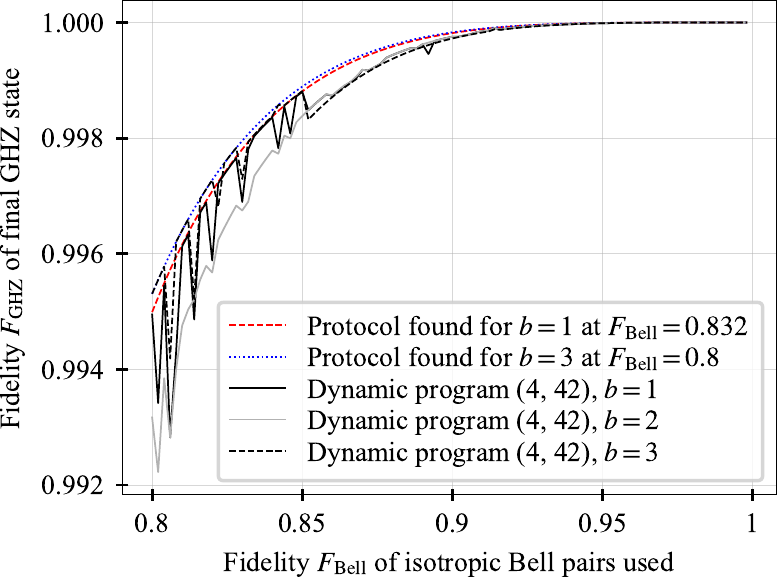}
\centering
\caption{Fidelity of the final 4-qubit GHZ as a function of the input Bell pair fidelity for a fixed number of input Bell pairs $k=42$. 
We plot (black solid line, gray solid line, and dashed black line) the results of the base dynamic program with a buffer size $b=1,2,3$ for each input Bell pair fidelity. We compare the results to the fidelity achieved by two fixed protocols for all input Bell pair fidelities (dashed red line and dotted blue line). 
The fixed protocols correspond with the protocols that the base dynamic program found for $b=1,F_\text{Bell}=0.832$ and $b=3,F_{\text{Bell}}=0.8$. They achieve approximately the convex hull of the individual protocols found by the base dynamic program with $b=1,b=3$ respectively. 
}
\label{fig:Plot1}
\end{figure}

We would like to stress that this is a heuristic approach and in general, it does not lead to the optimal algorithms. 
This can be observed in Fig. \ref{fig:Plot1}. We plot the fidelity of the produced $F_\text{GHZ}$ with respect to the fidelity $F_\text{Bell}$ of the input Bell pairs. The fidelity of the produced $F_\text{GHZ}$ is not always monotonically increasing. Moreover, as we increase the size of the buffer $b$ the program tends to find better protocols. However, lines with different values of $b$ cross highlighting the suboptimality of the output. 

\subsection{Randomized version of the dynamic program}
The base dynamic algorithm allows optimizing GHZ generation for moderate sizes. Unfortunately, even if faster than brute force, it still scales exponentially with the size of the GHZ state, and, for fixed GHZ size, it scales quadratically with the size of the buffer. In this section, we discuss a randomized version of the base dynamic program. This algorithm scales to larger GHZ sizes and, in practice, finds better protocols.

The randomized algorithm takes an additional parameter compared to the base dynamic algorithm. This is the temperature $T$, that is used to decide whether or not to keep intermediate protocols. The randomized algorithm has the same two loops over $n,k$ as the base one. It has an outer loop over the GHZ state size $n$ starting from $n=1$ to $n_{\max}$ and an inner loop over the number of Bell pairs starting from $k=n-1$ to $k_{\max}$. The two algorithms differ in how they construct the pool of protocols for $(n,k)$. 

Similar to the base algorithm, the randomized algorithm fills a pool of $b$ protocols for each combination of $(n,k)$. For $1\leq i\leq b$, the algorithm chooses between non-local measurement or fusion with probability one half. Then it selects uniformly at random a stabilizer to perform a non-local measurement or a fusion scheme to implement. Both the non-local measurement or the fusion scheme combine two smaller states, the parameters  $(n,k-k')$ and $(n',k')$ for the non-local measurement or $(n-n'+1,k-k')$ and $(n',k')$ for fusion are singled out by the scheme choice. Finally, the two states are chosen uniformly at random from the $b$ states stored. 

The $i$-th slot of the pool of $b$ states is filled with certainty with the new state if either 1) it is the first state---\textit{i.e.}, $i=1$---or 2) it achieves a higher fidelity than the previous protocol. If these conditions are not met, the new state can still probabilistically be accepted with probability $e^{\Delta F_\text{GHZ}/T}$, where $\Delta F_\text{GHZ}$ is the fidelity resulting from the new protocol minus the fidelity resulting from the previous protocol. If the value for $T$ is set at a high value, states with a lower fidelity than the fidelity of the previous protocol are more likely to be accepted. If the new protocol is not stored, the $i$-th slot is filled with the $i-1$-th state. This approach makes that protocols that lead to a high $F_\text{GHZ}$ end up in multiple buffer slots, and are therefore more likely to be randomly selected at larger values of $(n,k)$. 

\begin{figure}[]
\includegraphics[width=8.5cm]{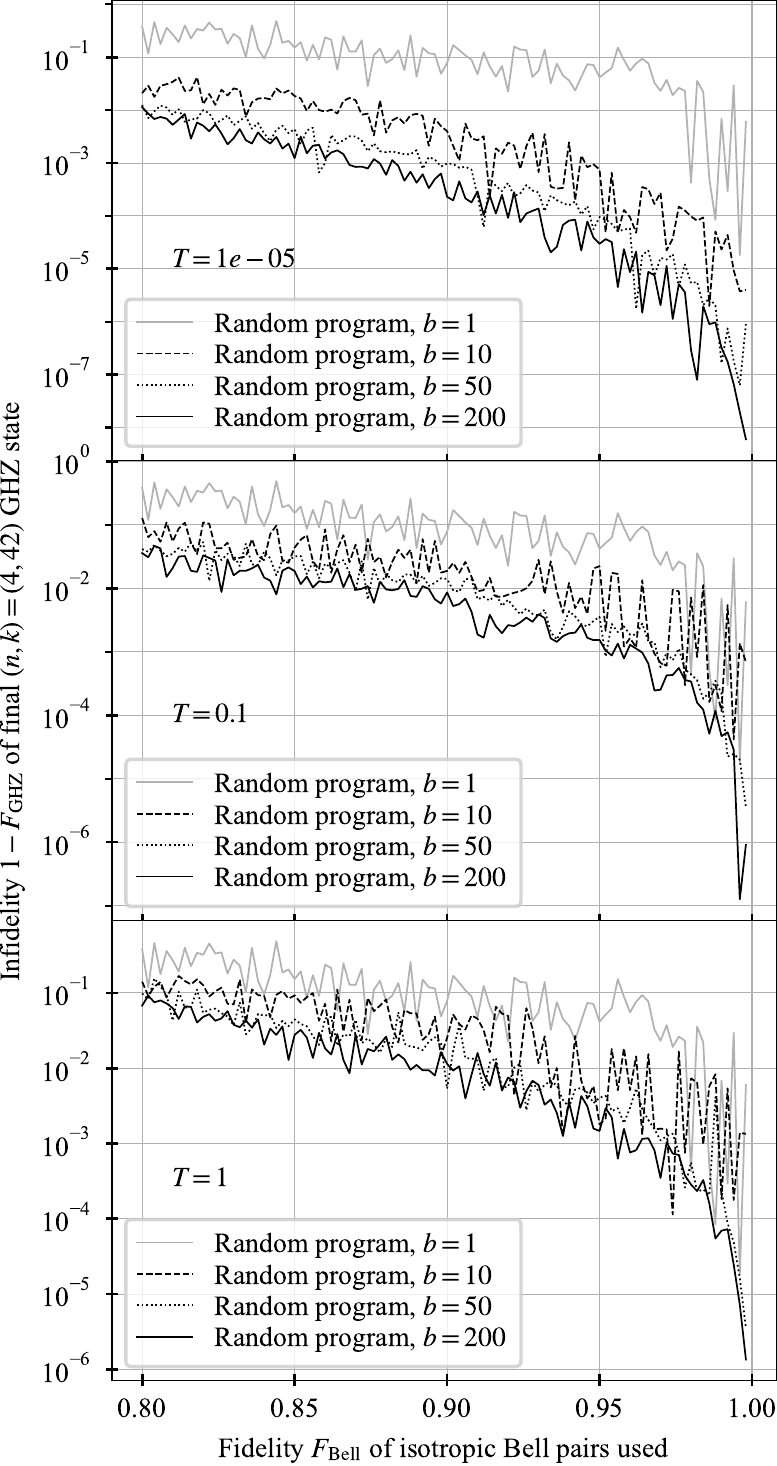}
\centering
\caption{Performance of the randomized dynamic program as a function of the configuration parameters, for different fixed temperatures; from top to bottom $T=0.00001$, $T=0.1$, and $T=1$.}
\label{fig:plot_rand_diff_temperatures_7616}
\end{figure}

In the following, we investigate the effect of the configuration parameters, the buffer size $b$ and the temperature $T$, on the performance of the algorithm. In particular, we fix $(n,k)=(4,42)$ and evaluate the fidelity of the final GHZ state as a function of the input Bell pair fidelity. We plot the results in three plots in \ref{fig:plot_rand_diff_temperatures_7616}, from top to bottom the temperatures are fixed to  $T=0.00001,T=0.1,T=1$. In each plot we show four lines corresponding to four different buffer sizes $b\in\{1,10,50,200\}$. We see that, in general, a lower temperature gives better results. This can be understood by realizing that for low temperatures, states with higher fidelity are more likely to be stored in many slots of the buffer. This also indicates that fidelity is a good measure for determining the quality of a protocol as a building block for a larger protocol.

\subsection{Comparison between the dynamic programs}
\begin{figure}[]
\includegraphics[width=8.5cm]{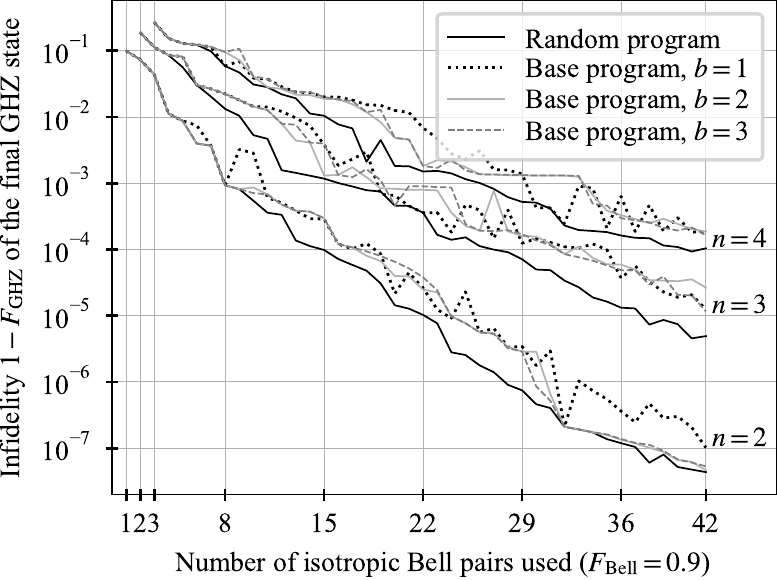}
\centering
\caption{Fidelity achieved by the protocols from the base and random dynamic programs as a function of the number of Bell pairs. The fidelity of the input Bell pairs is fixed to $F_\text{Bell}=0.9$. From top to bottom each set of four lines corresponds to GHZ size $n=4,3,2$. 
The lines labeled ``base program'' indicate the best protocols found by the base dynamic algorithm for different buffer sizes $b$. The black solid line indicates the best protocol found by the random variant. For the base algorithm, we used buffer sizes $b\in\{1,3\}$.
For the random algorithm, we ran the algorithm 44 times, with 18 different temperatures (between $T=0.00001$ and $T=0.0009$) per iteration and $b=200$.}
\label{fig:Plot2}
\end{figure}

We end the discussion of the dynamic programs by comparing the output of the base dynamic program against the randomized version. In Fig. \ref{fig:Plot2} we compare the best results with $(n,k)=(4,42)$ of the base dynamic program against the randomized program. For the parameters chosen, the randomized dynamic program outperforms the base dynamic program.

\section{Results} \label{sec:results}
In this section, we use the dynamic algorithms to find good GHZ creation protocols. First, we investigate how the different variants of the dynamic program discussed in section \ref{sec:dynamicprograms} compare with each other and what  the optimal parameter configurations are. Then, we use the programs to investigate scenarios of interest.  
First, given the importance of the surface code, we study the distribution of 4-qubit GHZ states. Second, we explore how the quality of the GHZ state for the best protocols scales with the number of parties $n$. 

The source code of the dynamic algorithm can be found online~\cite{deboneGHZProtSoftware2020}.

\subsection{Comparison with existing protocols for 4 parties}
\begin{figure}[]
\includegraphics[width=8.5cm]{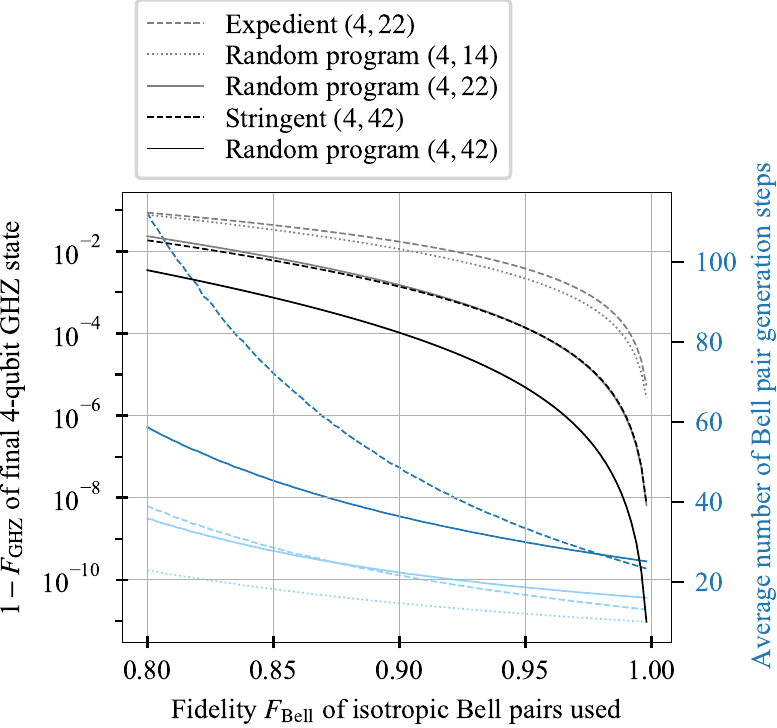}
\centering
\caption{Comparison between the \textit{Expedient} and \textit{Stringent} protocols~\cite{Nickerson2013a} (see Fig. \ref{fig:Expedient_Stringent}), and the best algorithms found with the dynamic programs for $(n,k)=(4,14)$, $(n,k)=(4,22)$ and $(n,k)=(4,42)$. The protocols are found with the randomized version of the dynamic program, using the settings and parameters discussed in Fig. \ref{fig:Plot2}. The blue lines show (right y-axis ticks) the average number of Bell pair generation steps for each of the protocols. To calculate this metric we take the creation of one Bell pair as one time step, and neglect the duration of all other elements of the protocols. The averages are calculated by executing the protocols 100000 times for each fidelity.}
\label{fig:Plot3}
\end{figure}

\begin{figure*}
\includegraphics[width=0.94\textwidth]{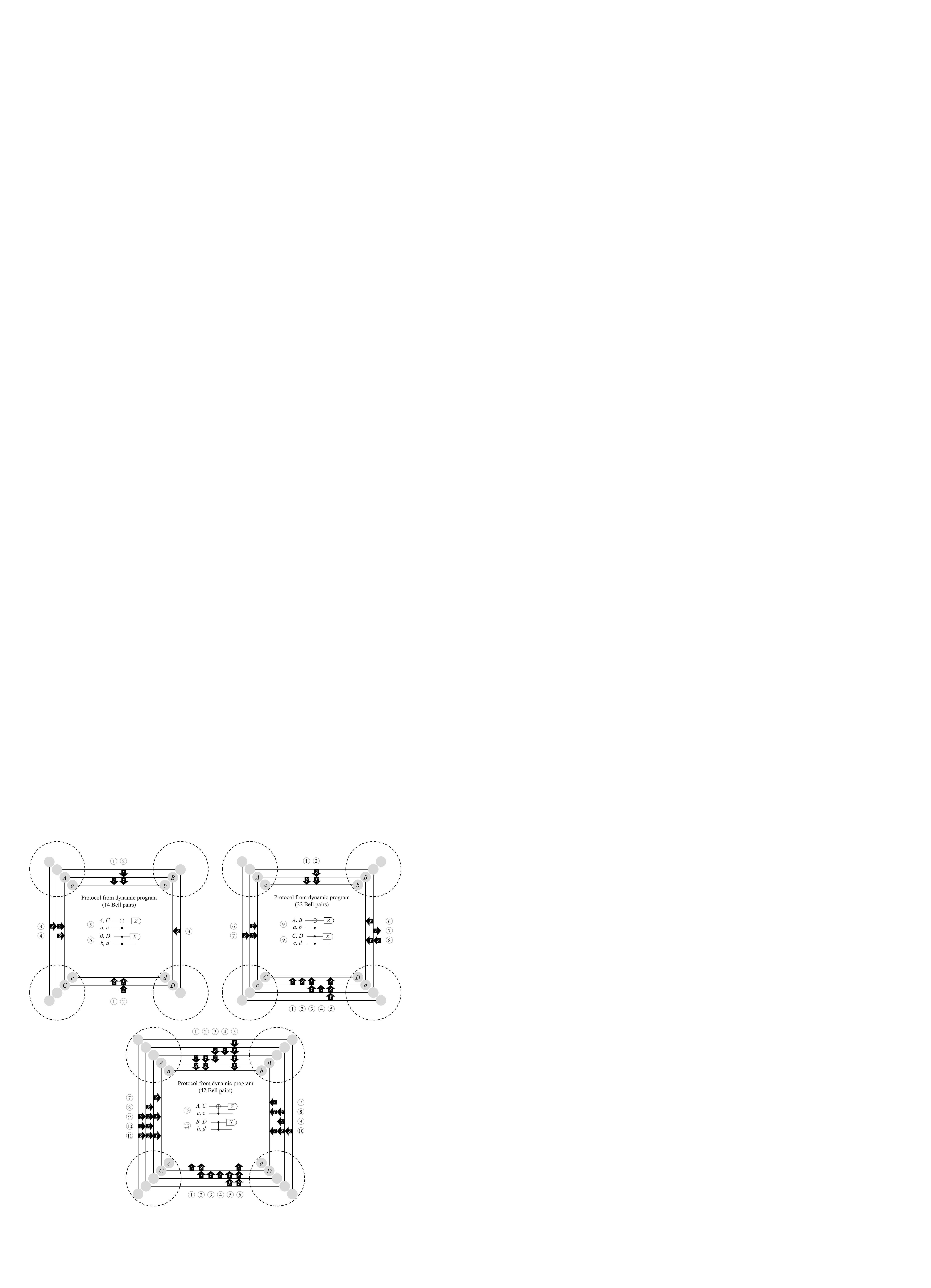}
\centering
\caption{Protocols for creating and purifying a 4-qubit GHZ diagonal state out of 14, 22 and 42 Bell diagonal states shared between 4 network parties, found with the randomized version of the dynamic program presented in this paper. See Fig. \ref{fig:Expedient_Stringent} for more information about the notation.}
\label{fig:dynprogprots}
\end{figure*}

First, we investigate protocols for $n=4$. In Fig. \ref{fig:Plot3} we compare the best protocols that our dynamic algorithms find for parameters $(n,k)=(4,14)$, $(n,k)=(4,22)$ and $(n,k)=(4,42)$ with the Expedient and Stringent protocols from Nickerson \textit{et al.} \cite{Nickerson2013a}. The figure shows the infidelity (one minus the fidelity) of the output GHZ state as a function of the fidelity of the input Bell pairs. We see that under the conditions considered here, the new protocols create higher quality GHZ states with the same or even with a smaller number of Bell states. On the other hand, while Expedient and Stringent require three qubits per node, these protocols typically require more qubits per node. For example, the $(n,k)=(4,14)$ protocol found with our dynamic program requires that two of the nodes have four qubits, as can be seen in Fig. \ref{fig:dynprogprots}. The best $(n,k)=(4,22)$ protocol found by the random dynamic program can be performed if all four nodes have four qubits (see Fig. \ref{fig:dynprogprots}), and the best $(n,k)=(4,42)$ found can be achieved with five qubits per node (see Fig. \ref{fig:dynprogprots}).

Let us now investigate to which degree the new protocols achieve these higher fidelities consuming a smaller amount of resources. We note that $k$ represents the \textit{minimum} number of Bell pairs needed to generate a GHZ state. These protocols are probabilistic, and the success probability depends on the fidelity of the states, going from one if the states are perfect to zero if the measured state is a minus one eigenstate of one of the measurement operators. If the protocol fails at some step, the step needs to be run again from the beginning. This means that the average number of Bell pairs might be very different from $k$.

\begin{figure}[b]
\includegraphics[width=8.5cm]{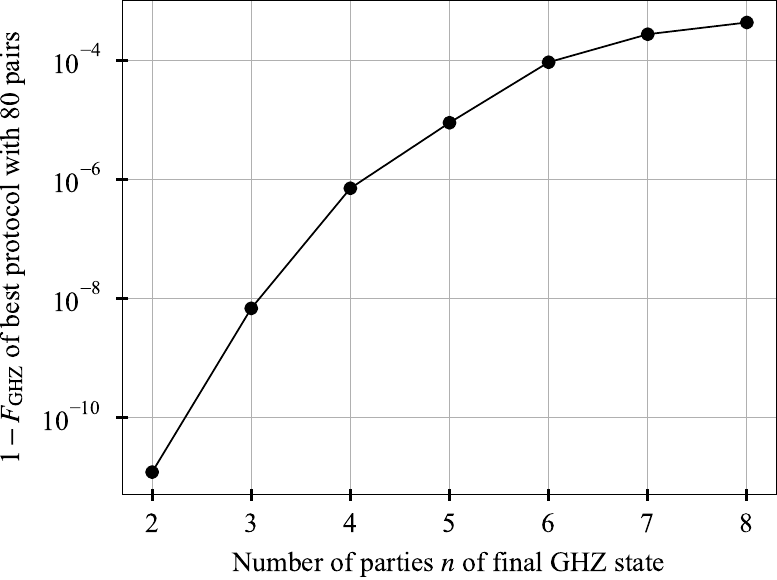}
\centering
\caption{Infidelity ($1-F_\text{GHZ}$) of the best protocols found for an $n$-qubit GHZ as a function of $n$ and input Bell pair fidelity $F_\text{Bell}=0.9$ and $k=80$. The protocols are found with the randomized version of the dynamic program, using the settings and parameters discussed in Fig. \ref{fig:Plot2}.}
\label{fig:Plot5}
\end{figure}

Another related figure of merit which might be more relevant in practice is the average number of entanglement generation steps. For this, we assume that network nodes can generate a Bell pair deterministically with one other node over some unit time step. Hence different pairs of nodes can generate entanglement in parallel over some unit time step, for instance, the two left nodes in Fig. \ref{fig:dynprogprots} can generate entanglement in parallel to the right nodes. If the duration of the time step is qualitatively larger than the gate time, the number of entanglement generation time steps represents to first-degree approximation the duration of the protocol.

We find that $k$ is a good proxy for the average number of entanglement generation steps. First, we see in Fig. \ref{fig:Plot3} that lower values of $k$ correspond with a lower average number of generation steps. Moreover, minimizing $k$ leads to a reduction in the average number of generation steps. Interestingly, the average number of generation steps of Stringent and Expedient cross with some of the new protocols. In particular, Stringent---a $(4,42)$ protocol---crosses with the new $(4,42)$ protocol, and  Expedient---a $(4,22)$ protocol---crosses with the new $(4,22)$ protocol. The reason for this is the higher symmetry in the structure of Expedient (Fig. \ref{fig:Expedient_Stringent}) and Stringent (Fig. \ref{fig:Expedient_Stringent}). 
In particular, they use the exact same number of Bell pairs at opposite sites of the network, whereas there are small differences in this respect for the new protocols. For very high input fidelities the success probabilities of all distillation steps are close to one which leads to a lower number of entanglement generation steps.

\begin{figure}[t]
\includegraphics[width=8.5cm]{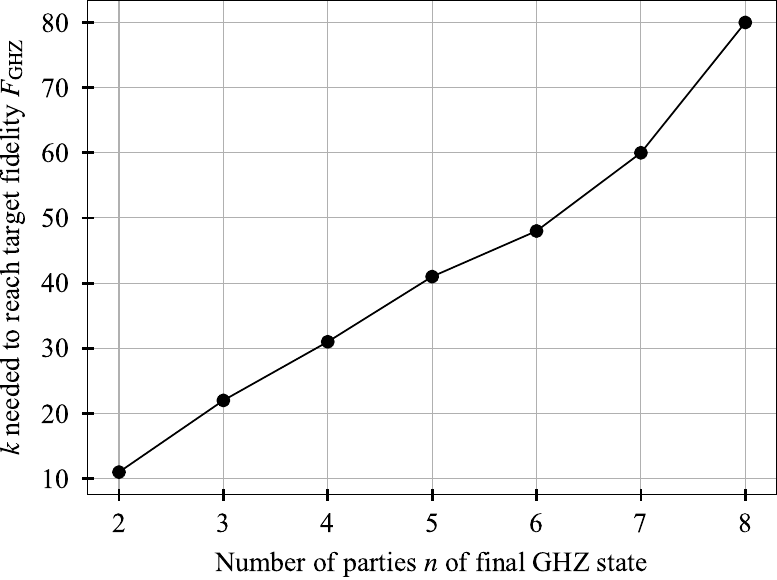}
\centering
\caption{Number $k$ of isotropic Bell pairs with $F_\text{Bell}=0.9$ needed to make an $n$-qubit GHZ state with $F_\text{GHZ}\geq 0.999565$ as a function of $n$. This is the best $F_\text{GHZ}$ found for an $(8,80)$ GHZ state with the random dynamic program in all our attempts. The protocols are found with the randomized version of the dynamic program, using the settings and parameters discussed in Fig. \ref{fig:Plot2}.}
\label{fig:Plot6}
\end{figure}

\subsection{Results for large number of parties}
Here, we investigate the trade-offs between the number of parties and the number of Bell pairs for $n>4$. 

First, for a fixed number of Bell pairs, we investigate how the GHZ fidelity drops as we increase the number of parties. In particular, we fix $k=80$ and vary $n$ from two to eight. In Fig. \ref{fig:Plot5} we show the fidelity $F_\text{GHZ}$ of the final $n$-qubit GHZ state as a function of $n$ for a fixed Bell pair fidelity $F_\text{Bell}=0.9$. We see that even for this high number of input pairs, the output fidelity drops sharply with the number of parties. 

Second, we invert the question and we investigate how many Bell pairs are necessary to achieve a fixed target GHZ fidelity $F_{\text{GHZ}}=0.999565$ for a different target number of parties. This is the best $F_\text{GHZ}$ found for an $(8,80)$ GHZ state with the random dynamic program in all our attempts. In Fig. \ref{fig:Plot6} we show the number of Bell pairs $k$ with fidelity $F_\text{Bell}=0.9$ needed to create an $n$-qubit GHZ state with fidelity $F_\text{GHZ}\geq 0.999565$. We observe that for the available data points the number of pairs scales roughly linearly with the number of parties.

\section{Conclusions} \label{sec:conclusions}
In this paper, we searched for protocols that generate high-fidelity GHZ states out of non-perfect Bell states. The goal was to minimize the number of Bell pairs for creating a high-quality GHZ state. We did this by using a dynamic program to search the protocol space, allowing two types of operations: fusion and distillation.

We found protocols that distill GHZ states with higher fidelity compared to previously known protocols. Compared to previous research \cite{Nickerson2013a}, the protocols found require roughly half the number of pairs to achieve a similar fidelity. Out of the different algorithm variants that we implemented, the randomized version found the best protocols in most regimes. For $n=2$ to $n=8$ parties involved, we investigated how the fidelity of the final GHZ state decreased by increasing the number of parties with a fixed number of pairs ($k=80$), and calculated how many Bell pairs are needed for the distribution of an $n$-qubit GHZ state of a fixed fidelity. Our programs can be used to find protocols for an arbitrary number of parties and entangled states involved.

GHZ states are required for implementing error-correction codes in distributed quantum computing. The distributed implementation of codes beyond the surface code has not been thoroughly explored. 
However, the case for the surface code in distributed implementations is weaker \cite{campbellRoadsFaulttolerantUniversal2017}. 
Richer connectivities than direct neighbors can be achieved with relative ease. 
Our tools open the door to implementations of alternative quantum error correcting codes that require high-quality GHZ states of weights different than four. 

While these results are promising, future work should quantify the precise effect on the noise threshold for distributed implementations of the surface code. For this, a starting step will be to evaluate how the new protocols behave in more realistic scenarios. In particular, it would be interesting to perform a follow-up of the search for GHZ protocols including noise and loss.

\section*{Acknowledgements}
The authors would like to thank Tim Coopmans, Stacey Jeffery, Tim Taminiau and Filip Rozp\k{e}dek for helpful discussions and feedback. This work was supported by the Netherlands Organization for Scientific Research (NWO/OCW), as part of the Quantum Software Consortium program (project number 024.003.037 / 3368). R. Ouyang was supported by Tsinghua Xuetang Talents Program.

\end{document}